\documentclass[12pt,preprint]{aastex}

\shorttitle{The 2006 Outburst of PKS\,2155$-$304}
\shortauthors{Sakamoto et al.}

\begin{document}

\title{CANGAROO-III Observations of the 2006 Outburst of PKS\,2155$-$304}

\author{
Y.~Sakamoto\altaffilmark{1},
K.~Nishijima\altaffilmark{2},
T.~Mizukami\altaffilmark{3},
E.~Yamazaki\altaffilmark{2},
J.~Kushida\altaffilmark{2},
R.~Enomoto\altaffilmark{4},
M.~Ohishi\altaffilmark{4},
G.V.~Bicknell\altaffilmark{5},
R.W.~Clay\altaffilmark{6},
P.G.~Edwards\altaffilmark{7},
S.~Gunji\altaffilmark{8},
S.~Hara\altaffilmark{9},
T.~Hattori\altaffilmark{2},
S.~Hayashi\altaffilmark{10},
Y.~Higashi\altaffilmark{3},
Y.~Hirai\altaffilmark{11},
K.~Inoue\altaffilmark{8},
C.~Itoh\altaffilmark{12},
S.~Kabuki\altaffilmark{3},
F.~Kajino\altaffilmark{10},
H.~Katagiri\altaffilmark{13},
A.~Kawachi\altaffilmark{2},
T.~Kifune\altaffilmark{4},
R.~Kiuchi\altaffilmark{4},
H.~Kubo\altaffilmark{3},
R.~Mizuniwa\altaffilmark{2},
M.~Mori\altaffilmark{4},
H.~Muraishi\altaffilmark{14},
T.~Naito\altaffilmark{15},
T.~Nakamori\altaffilmark{3},
S.~Nakano\altaffilmark{3},
D.~Nishida\altaffilmark{3},
A.~Seki\altaffilmark{2},
V.~Stamatescu\altaffilmark{6},
T.~Suzuki\altaffilmark{11},
D.L.~Swaby\altaffilmark{6},
T.~Tanimori\altaffilmark{3},
G.~Thornton\altaffilmark{6},
F.~Tokanai\altaffilmark{8},
K.~Tsuchiya\altaffilmark{3},
S.~Watanabe\altaffilmark{3},
Y.~Yamada\altaffilmark{10},
S.~Yanagita\altaffilmark{11},
T.~Yoshida\altaffilmark{11},
T.~Yoshikoshi\altaffilmark{4},
and Y.~Yukawa\altaffilmark{4}
}

\email{kyoshi@tkikam.sp.u-tokai.ac.jp}

\altaffiltext{1}{Graduate School of Science and Technology,
Tokai University, Hiratsuka, Kanagawa 259-1292, Japan}
\altaffiltext{2}{Department of Physics, \ 
Tokai University, Hiratsuka, Kanagawa 259-1292, Japan}
\altaffiltext{3}{Department of Physics, Graduate School of Science, 
Kyoto University, Sakyo-ku, Kyoto 606-8502, Japan}
\altaffiltext{4}{Institute for Cosmic Ray Research, 
University of Tokyo,  Kashiwa, Chiba 277-8582, Japan}
\altaffiltext{5}{Research School of Astronomy and Astrophysics,
Australian National University, ACT 2611, Australia} 
\altaffiltext{6}{Department of Physics and Mathematical Physics, 
University of Adelaide, SA 5005, Australia}
\altaffiltext{7}{Paul Wild Observatory, CSIRO Australia Telescope
National Facility, Locked Bag 194, Narrabri, NSW 2390, Australia}
\altaffiltext{8}{Department of Physics, Yamagata University, 
Yamagata, Yamagata 990-8560, Japan}
\altaffiltext{9}{Ibaraki Prefectural University of Health Sciences, 
Ami, Ibaraki 300-0394, Japan}
\altaffiltext{10}{Department of Physics, Konan University, 
Kobe, Hyogo 658-8501, Japan}
\altaffiltext{11}{Faculty of Science, Ibaraki University, 
Mito, Ibaraki 310-8512, Japan}
\altaffiltext{12}{National Institute of Radiological Sciences,
Chiba, Chiba 263-8555, Japan}
\altaffiltext{13}{Department of Physical Science, Graduate School 
of Science, Hiroshima University, Higashi-Hiroshima, 
Hiroshima 739-8526, Japan}
\altaffiltext{14}{School of Allied Health Sciences, 
Kitasato University, Sagamihara, Kanagawa 228-8555, Japan}
\altaffiltext{15}{Faculty of Management Information, 
Yamanashi Gakuin University, Kofu, Yamanashi 400-8575, Japan}

\begin{abstract}
We have used the CANGAROO-III imaging atmospheric Cherenkov telescopes 
to observe the high-frequency-peaked BL Lacertae (HBL) object
PKS\,2155$-$304 between 2006 July 28 (MJD 53944) and August 2,
triggered by the H.E.S.S.\ report that the source was in a high 
state of TeV gamma-ray emission.
A signal was detected at the $4.8~\sigma$ level
in an effective live time of 25.1\,hours 
during the outburst period.
The flux of Very High Energy gamma-rays from the CANGAROO-III observations
shows the variability on the time scale of less than 
a few hours.
The averaged integral flux above 660~GeV is
$(1.6\pm0.3_{stat}\pm0.5_{syst})\times10^{-11}\,\mbox{cm}^{-2}\,\mbox{sec}^{-1}$
which corresponds to $\sim45\%$ of the flux observed 
from the Crab nebula. 
Follow-up observations between August 17 (MJD 53964) and 25 
indicate the source activity had decreased.
\end{abstract}


%------------------------------------------------------------------------------
\section{Introduction}

The most astonishing contribution from the ground-based IACTs
has been the detection of AGNs at energies above several hundred GeV.
To date, 19 active galactic nuclei have been reported to emit 
Very High Energy (VHE) gamma-rays \citep{Hin07}, 
most of them classified as
high-frequency-peaked BL Lacs (HBLs) \citep{Pad95}.
Some, such as Mrk~421 and Mrk~501, have been targets of 
simultaneous multi-wavelength campaigns since their first
TeV detection \citep{Pun92, Qui96}, and are well studied.
The derived broadband spectral energy distributions (SEDs)
show that they have two components. The lower energy component
is believed to be synchrotron emission produced 
by relativistic electrons in the jet, and the higher energy 
component, although relatively poorly understood,  
is probably explained by the inverse-Compton scattering
of seed photons (either synchrotron or ambient photons)
by the same population of electrons 
\citep[see e.g.,][]{Jon74, Ulr97, Der93, Sik94}.

Multi-wavelength observations of HBLs have also shown that
the fluxes are extremely variable, particularly
in the higher energy bands.
Flux variations on a wide range of time scales
from months to two minutes
are reported at TeV energies 
\citep[e.g.,][]{Gai96, Kra01, Aha02, Aha05c, Bla05, Aha99, Gli06, Alb07a}.
Time variability is a very useful way to probe the emission mechanism
in the jet.
Characteristic time scales, which can be employed as a measure of
super-massive black hole mass at the center of HBL,
have been studied with X-ray data using statistical techniques
such as power spectra, structure functions, {\it etc\/} 
\citep[e.g.,][]{Hay98, Zha99, Kat01, Cui04}.
A good correlation of the X-ray and the TeV fluxes
has also been reported 
\citep[e.g.,][]{Buc96, Tak96, Dja99, Tak00, Sam00, Kra02, Alb07a}.
Recently, however, poor correlations or a total lack of correlation 
between the X-ray and TeV fluxes of individual flares
have been observed
\citep[e.g.,][]{Bla05,Aha05c}, 
and even TeV flares with no X-ray counterpart have been found 
\citep{Bla05, Kra04}.  
Spectral hardening with flux in the TeV energy range 
has been reported for 
Mrk~421 \citep{Aha02, Kre02, Alb07a} and Mrk~501 \citep{Alb07b}. 

PKS\,2155$-$304 ($z=0.116$ \citep{Fal93}) is one of the brightest BL Lacs 
in the X-ray \citep{Bri94,Kub98,Gio98,Ves99,Nic02,Zha05} 
and EUV \citep{Mar93} bands. 
Since the discovery of X-ray emission from 
this object \citep{Gri79,Sch79}, 
it has been repeatedly observed
over a wide range of frequencies from radio to 
Very High Energy (VHE) gamma-rays
\citep[e.g.,][]{Tre89,Ves95,Zha96,Pia97,Pin04,Dom04,Aha05a}.

In 1997 November, a gamma-ray and X-ray outburst
from PKS\,2155$-$304 was detected by EGRET \citep{Sre97},
BeppoSAX \citep{Chi99} and RXTE \citep{Ves99}.
During this active phase, the Durham group observed
PKS\,2155$-$304 using their Mark~6 telescope,
and reported the first detection of VHE gamma-rays
at the 6.8\,$\sigma$ level above 300\,GeV \citep{Cha99a} 
at the time of BeppoSAX observation.
However they found no evidence for TeV gamma-ray emission during
observations in 1998 \citep{Cha99b} when the X-ray flux level was low.
PKS\,2155$-$304 was observed with
the CANGAROO-I 3.8\,m telescope in 1997, though poor weather resulted in
minimal overlap with the Durham observations.
No gamma-ray signal above 1.5\,TeV was detected \citep{Rob99}.
PKS\,2155$-$304 was further observed in 1999, 2000 and 2001
with the CANGAROO-II telescope.
It remained in a low state of X-ray activity 
in those periods,
and was not detected above the energy threshold
of 420\,GeV \citep{Nis01,Nis02,Nak03}.
PKS\,2155$-$304 was confirmed as a TeV gamma-ray source
by the H.E.S.S.\ group in 2004. They reported a 45\,$\sigma$
detection 
at energies greater than 160\,GeV in July and October, 2002,
and June--September, 2003 \citep{Aha05a}. 
The flux variability on time scales 
of months, days, and hours were also reported, 
and the monthly-averaged flux above 300\,GeV 
was between 10\% and 60\% of Crab flux. 
The energy spectrum was characterized by a steep power law 
with a time-averaged photon index of $\Gamma\sim3.3$.
Multi-wavelength observations in 2003 in a low state showed
no correlation between the X-ray and the gamma-ray fluxes,
or between any other wavebands, even though the fastest ever 
X-ray flare in this object, with a $1500$~sec timescale, 
was detected \citep{Aha05b}. 

In July 2006, the H.E.S.S.\ group reported that PKS\,2155$-$304
had been detected at historically high TeV flux levels of up 
to several crab
\citep{Ben06,
%\footnote{ATel 867 is available at:
%http://www.astronomerstelegram.org/?read=867}, 
Rau06}.
They also presented very rapid flux variability 
with one-minute time scale resolution 
and found well-resolved bursts varying on timescales of
$\sim200$ seconds \citep{Aha07}. 
The H.E.S.S.\ report triggered  multi-wavelength 
Target of Opportunity (ToO) observations, 
including the observations with the CANGAROO-III telescopes 
from 2006 July 28 (MJD 53944) described in this paper.
%\citet{Fos07} report the X-ray flux increased by a factor of 5
%and the optical/UV flux by a factor of 1.5, 
%with a delay of roughly 1 day.

Numerous multi-wavelength campaigns of PKS\,2155$-$304
have been carried out
\citep[e.g.,][]{Ede95,Urr97,Aha05b},
and many physical implications for the emission mechanisms have been
reported \citep[e.g.,][]{Chi99,Kat00,Ede01,Tan01,Zha02,Zha06}.
However, these models are still face challenges in
explaining the complex and different patterns of 
multi-wavelength variations for each epoch. 

For objects like PKS\,2155$-$304 which show rapid and 
complex time variability,
continuous monitoring is very important for modeling
the emission mechanisms.     
The difference in longitude between 
the H.E.S.S.\ and CANGAROO-III 
sites is $\sim120^{\circ}$, corresponding to an
8\,hour time difference. Thus the H.E.S.S.\ 
and the CANGAROO-III data complement one another.
Northern hemisphere blazars, such as Mrk~421, Mrk~501 and 
1ES~1959$+$650, have been observed at TeV energies
continuously with more than two geographically distant 
telescope systems
\citep[e.g.,][]{Reb06,Gli06,Kra04}, 
providing greatly improved time coverage, but
PKS~2155$-$304 is the first object in the southern sky
for which such studies have been made.
 
%------------------------------------------------------------------------------
\section{Detector and Observations}

The CANGAROO-III imaging atmospheric Cherenkov telescope
system which consists of four telescopes are
operated in Woomera, South Australia (longitude $136^{\circ}47'$\,E,
latitude $31^{\circ}06'$\,S, 160~m~a.s.l.).
They have parabolic reflectors of 10~m~diameter
with an 8~m~focal length. Each reflector 
consists of 114 small spherical mirrors
of 80~cm~diameter, which are made of Fiber Reinforced Plastic.
The imaging cameras have 427 pixels of $0.17^{\circ}$ size and
a total field of view of $\sim4^{\circ}$.
These telescopes use altitude-azimuth mounts and are placed at the corners
of a diamond with 100~m~sides. 
The details of the mirror, the camera, 
and the total performance of the light collecting system 
are described in \citet{Kaw01}, \citet{Kab03}, and \citet{Eno06a}.

The CANGAROO-III observations of PKS\,2155$-$304 
during this outburst period were made
on five moonless nights between July 28 (MJD 53944) to August 2 in 2006.
There were no observations on August 1, and the July 29 observations were
affected by cloud.
Three (T2, T3, T4) of the four telescopes were used in these observations.
For each telescope, when the number of hit pixels 
exceeds four within a 20~nsec coincidence window, a local trigger
is sent to the stereo trigger system through an optical link.
In the stereo trigger system, determination of stereo events 
is done by requiring that at least two local trigger signals 
coinciding for at least 10~nsec within 
a 650~nsec time window, taking the geometrical time delay,
% considering the geometrical time delay 
%which depends on the telescope pointing direction \citep{Kub01,Nis05}. 
which depends on the telescope pointing direction,
into consideration\citep{Kub01,Nis05}. 

These observations were made using wobble mode
in which pointing position of each telescope was shifted in declination
between $\pm0.5^{\circ}$ from the center of PKS\,2155$-$304 alternatively
every twenty minutes.
Due to a mechanical tracking problem with the T3 telescope during
this period, stereo observations
with three telescopes were done only after culmination.
Stereoscopic observations using T2 and T4, which are located on 
the long diagonal of the diamond,
were performed before culmination from July 28 to July 31.

The details of the CANGAROO-III observations of 
PKS\,2155-304 are summarized in Table~\ref{mjd}.
The typical trigger rates, the mean zenith angle,
and the total observation time of two-fold coincidences of T2--T4
were $\sim$20~Hz, $26.6^{\circ}$, and 11.4\,hours, respectively.
The typical trigger rate of three-fold coincidences
was $\sim$12~Hz, with a mean zenith angle for these
observations of $20.4^{\circ}$
and a total observation time of 17.6\,hours.
The daily observation time $t_{obs}$, average zenith angle $z$,
and trigger rate $r_{tr}$ are shown 
in Table~\ref{mjd} for both T2--T4
and three-fold data.
The zenith angles of observations ranged from 
$52^{\circ}$ to $7^{\circ}$ for T2--T4 data, and from
$1^{\circ}$ to $46^{\circ}$ for three-fold data.

Follow-up observations were made with the same system on six moonless
nights between August 17 (MJD 53964) and 25.  
These data were taken only after
culmination using three-fold coincidences.  The zenith angle during these
observations ranged from $1^{\circ}$ to $48^{\circ}$ with a mean of
$20.9^{\circ}$, and a total observation time of 19.1\,hours.
These observations are also summarized 
in Table~\ref{mjd}.

%------------------------------------------------------------------------------
\section{Analysis}

In order to calibrate the relative gain and the timing
for each pixel, the data taken by illuminating the focal plane 
uniformly with LED photons were used.
Conversion factors from ADC value to photoelectrons
for each pixel are determined using the LED system 
in the camera vessel,  
and the timing jitter of the pulse signals are
calibrated using light from a LED mounted at the center of
the telescope dish. 
Details of the standard CANGAROO-III calibration
methods are given in \citet{Kab03}.
These LED runs were performed at least once per night.
Five out of 427 pixels for T2 and two for T3
were identified as bad pixels which gave no signal or 
false signals, and these were removed from the analysis. 
There were no bad pixels in the T4 camera. 

\begin{figure}
\epsscale{}
\plotone{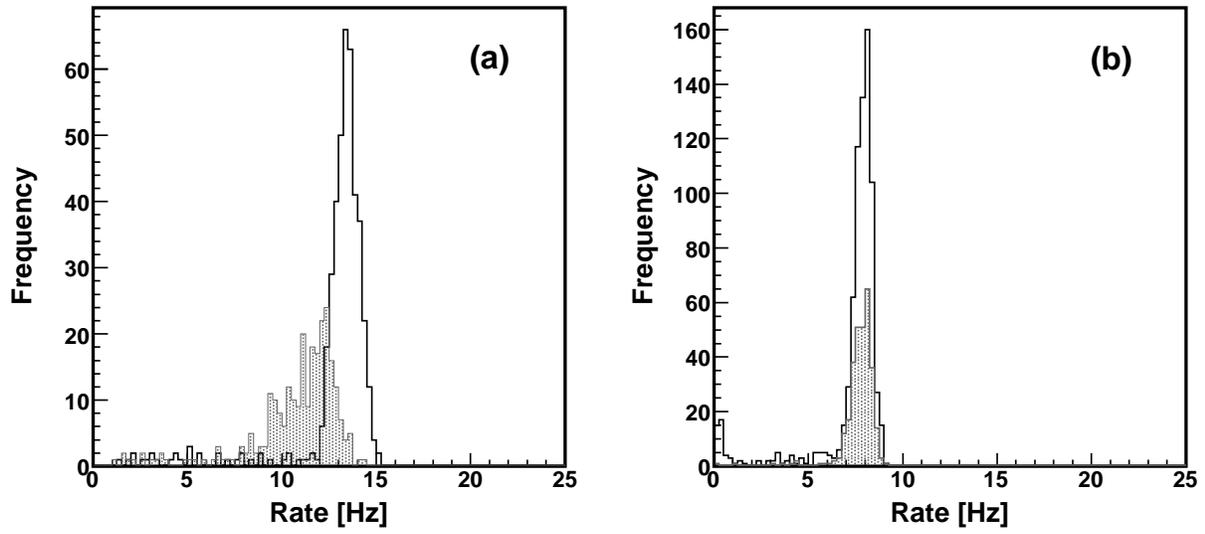}
\caption{Distributions of shower rate after image cleaning 
for $2fold$(left panel) and $3fold$(right panel).
Solid and hatched histograms show ones for zenith angles
less than $30^{\circ}$ and larger than $30^{\circ}$, respectively.
From our analysis we exclude the data whose shower rate are 
lower than 10~Hz, 8~Hz, and 5~Hz for $2fold$--$SZ$, $2fold$--$LZ$, 
and $3fold$, respectively.
\label{shower_rate}}
\end{figure}

We have cleaned the images to eliminate
night sky background photons and select shower images.
Each pixel is required to have a signal larger than 5.0\,p.e.\
with an arrival time within
$\pm$30\,nsec of the average shower arrival time.
Images containing at least five adjacent pixels
which pass the above requirements are selected.
The distributions of shower rate each minute are
shown in Fig.~\ref{shower_rate} for T2--T4 coincidence
(left panel) and three-fold coincidence(right panel), 
respectively, and
the average shower rate $r_{sh}$ for each night 
is listed in Table~\ref{summary}.
In order to ensure reliable arrival direction and 
energy estimations, the data taken during periods 
when shower rate was lower than $\sim$5~Hz
for three-fold coincidences and $\sim$10~Hz for T2--T4 coincidences
are not used, as they are probably affected by clouds.
In the following analysis, we further require 
that none of the brightest 15 pixels of each image
should be in the outermost layer of the camera 
in order to avoid deformation of the image. 
This method improves the effective area 
by more than 20\% compared to the previous simple edge cut \citep{Eno06a},
particularly above $\sim$2~TeV.
After taking into account the DAQ dead-time, 
the effective total live time $t_{liv}$ is also summarized 
in the same table.
 
The moments of the shower image are then parameterized
using the {\it Hillas parameters\/} \citep{Hil85},
and the arrival directions are reconstructed
using the intersection of image axes.
The intersection point is obtained by minimizing 
the sum of squared widths of the images
seen from the assumed point with a constraint on the distances
between images' center of gravity and assumed intersection point (IP-fit)
considering the $length$/$width$ ratio (see \citet{Kab07} for details),
which is similar to Algorithm 5 in \citet{Hof99}.
This method improves the signal to noise ratio by more than 10\%
compared to the procedure described in \citet{Eno06a} and
is confirmed by Monte Carlo simulations and observations of the Crab nebula.

After event reconstruction, numerous cosmic ray 
background events are rejected using
the Fisher Discriminant (FD) method \citep{Fis36}.
The FD value for each event is calculated as explained
in \citet{Eno06b} using the size-corrected $width$ and $length$ 
for each telescope, where the energy dependence of 
the $width$ and $length$ are corrected to keep each parameter 
at the same values in various energies. 
In order to determine the optimum FD cut values, 
a Monte Carlo simulation was carried out 
assuming a single power-law spectrum, $dF/dE \propto{E^{-\Gamma}}$,
with an index of $\Gamma=3.3$,
as determined for PKS~2155$-$304
by the H.E.S.S.\ group \citep{Aha05a}.
Before applying the FD cuts, the data are divided into two data sets,
''$2fold$'' which include the data taken by only T2--T4 coincidence
and ''$3fold$'' which required a three-fold coincidence between the 
three telescopes.
Each dataset is further divided by zenith angle into two datasets,
''$3fold$--$SZ$'' and ''$2fold$--$SZ$'' for $z<30^{\circ}$, and 
''$3fold$--$LZ$'' and ''$2fold$--$LZ$'' for $z>30^{\circ}$, respectively. 
Considering the Point Spread Function (PSF) of $0.23^{\circ}$ (FWHM),
the ${\theta}^2$ (where $\theta$ is the angular difference
between the reconstructed arrival direction and the source position) 
cuts were applied at $\theta^{2}<0.06~\mbox{deg}^2$ 
for the $3fold$ data, 
and to keep the same efficiency for gamma rays as for 
the three-fold analysis,   
$\theta^{2}<0.14~\mbox{deg}^2$ was applied to the $2fold$ data.
Based on the Monte Carlo simulation assuming 100~\% Crab flux 
of gamma rays above 600~GeV, 
we could determine the best FD cut criteria 
which yielded the highest significance for each dataset.
Following the results from the Monte Carlo study,
we select the events with $FD>-0.3$ for $3fold$--$SZ$, 
$FD>-0.2$ for $3fold$--$LZ$, $FD>-0.1$ for $2fold$--$SZ$
and $FD>-0.2$ for $2fold$--$LZ$ datasets, as candidate gamma-ray events.

The primary gamma-ray energy is estimated from the number of
detected photo-electrons, based on the Monte Carlo simulations
assuming a single power-law spectrum as mentioned above.
This relation depends on the zenith angle of observations,
so the simulations were done using the same variation of elevation 
as the actual observations. 
Although there is also some dependence on the impact parameter in the
energy determination, we do not incorporate that here, with the
resulting energy resolution estimated to be 30~\% around 1 TeV.
The effective detection area has been estimated 
from the Monte Carlo simulation as a function of energy.
The detection energy threshold $E_{th}$ is taken to be
the energy of the peak of the distribution of triggered shower energies.
                              
%------------------------------------------------------------------------------
\section{Results}

In this section, we show the results of CANGAROO-III observations
of PKS\,2155$-$304 in 2006. A summary of the results are shown 
in Table~\ref{summary}. Using these results, we present
the average integral flux, the average differential 
energy spectrum, and the time variation.

\subsection{Average flux}

\begin{figure}
\epsscale{}
 \plotone{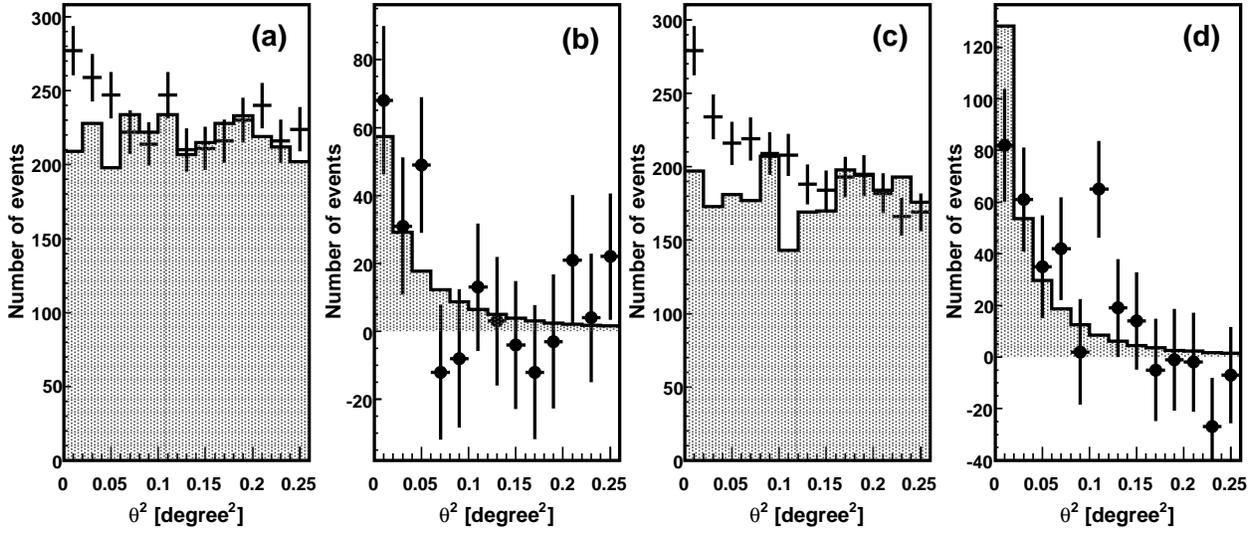}
\caption{Distributions of squared angular distances $\theta^2$ of
$2fold$ and $3fold$ data for PKS\,2155$-$304 in the outburst period,
obtained after the FD cut.
(a):On-source data (error bars) and off-source data (hatched histogram)
for $2fold$ are overlaid, where the latter is normalized to the live time 
(b):Closed circles show the excess events of the on-source above 
the off-source level for $2fold$. The hatched histogram represents
a PSF normalized to the number of excess events in $\theta^2<0.25$.
(c):Same as (a) for $3fold$.
(d):Same as (b) for $3fold$.
\label{theta2}}
\end{figure}

\begin{figure}
\epsscale{}
\plotone{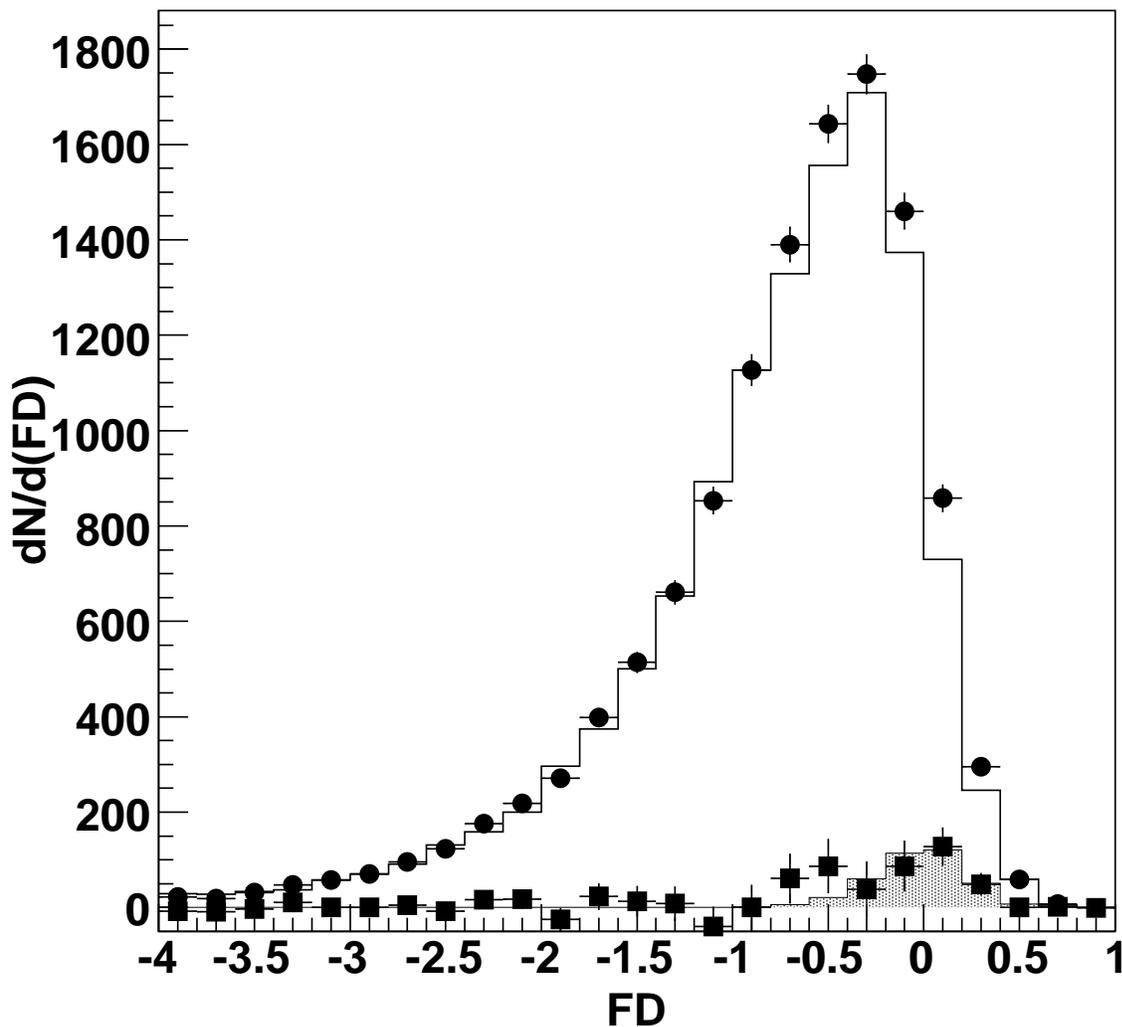}
\caption{Fisher Discriminant (FD) distribution for all events
after $\theta^{2}$ cut in the outburst period.
Closed circles with error bars are on-source
events ($\theta^2<0.06$ for $3fold$ data and $\theta^2<0.14$ for
 $2fold$),
and a solid histogram indicates off-source events.
The closed squares show the distribution of on$-$off events 
which are gamma-ray candidates, and a dashed histogram is 
an expected distribution from Monte Carlo gamma-ray events
normalized to the number of excess events.
\label{FD}}
\end{figure}

After the data reduction described in previous section, the final
${\theta}^2$ distribution
is obtained.
Fig.~\ref{theta2} (a) and (c) show the results in the outburst period 
for $3fold$ and $2fold$ datasets, respectively.
The background level was estimated using off-source data
in the corresponding region in the other side of the field of view.
The distributions of the excess events against $\theta^{2}$,
i.e., with the off-source events subtracted from the on-source events,
are shown in Fig.~\ref{theta2}(b) and (d) for $3fold$ and $2fold$, 
respectively.
There are clear excesses at small values of ${\theta}^2$ 
in both figures corresponding to the observed signal 
from PKS\,2155$-$304. 
The number of excess events are calculated using a circular region 
of $\theta^{2}<0.06~\mbox{deg}^2$ for $3fold$ and 
$\theta^{2}<0.14~\mbox{deg}^2$ for $2fold$ 
centered on the source position, respectively,
considering the PSF as mentioned in the previous section.
The hatched histograms in figure~\ref{theta2} (b) and (d) 
indicate the expected PSF normalized to the number of
excess events in $\theta^{2}<0.25$.
The combined excess of all the data is 
322$\pm$67~events ($4.8\sigma$), 
where only the propagation of statistical errors is considered.
Fig.~\ref{FD} shows the FD distributions after the $\theta^2$ cut
for on-source events, off-source events and the gamma-ray 
candidate events (on-source$-$off-source) together with
the Monte-Carlo gamma-rays normalized to the number of 
excess events.
The FD distributions are quite consistent for
the observed gamma-ray events and the Monte Carlo gamma rays.  
The FD cut is not applied for this plot.
The time-averaged integral flux above 660~GeV is
calculated to be 
$F(>~660~\mbox{GeV})~=~(1.6\pm0.3_{stat}\pm0.5_{syst})\times10^{-11}~\mbox{cm}^{-2}~\mbox{sec}^{-1}$.
This corresponds to $\sim45\%$ of the flux observed 
from the Crab nebula \citep{Aha04}, and is a factor of five
more intense than the flux in the low state reported in \citep{Aha05a}.
The systematic error in the flux arises mainly from 
energy scale uncertainties due to the absolute 
light collection efficiency (20\%), FD cut criteria (22\%),
uncertainties in the probability density function of images 
generated by Monte Carlo simulation (10\%), and the uncertainty of 
a power law index (8\%).
At a higher energy threshold above 1.0 TeV, which corresponds
to the energy threshold for $3fold$--$LZ$ and $2fold$--$LZ$ data,
no significant excess above our sensitivity from PKS\,2155$-$304 
is found in this period, 
and the resulting $2\sigma$ flux upper limit is
$F(>~1.0~\mbox{TeV})<9.2\times10^{-12}~\mbox{cm}^{-2}~\mbox{sec}^{-1}$.

For a cross-check, an alternative analysis using
the Fisher Discriminant with a fit \citep{Eno06b} 
was performed on the same data.
Although this is our standard method of analysis,
the fine time binning and consequent small count statistics
make it difficult to estimate the systematic errors correctly.
The main source of error is the non-uniformity of 
the Fisher Discriminant in the ring-shaped background region.
So we primarily took a conservative FD cuts method 
using a single reflected background at the expense 
of the statistics, and used FD fits 
as a cross-check. 
The difference of the integral flux between them
was less than $15\%$ comparing at the same threshold energy.

\subsection{Average differential energy spectrum}

\begin{figure}
\epsscale{}
\plotone{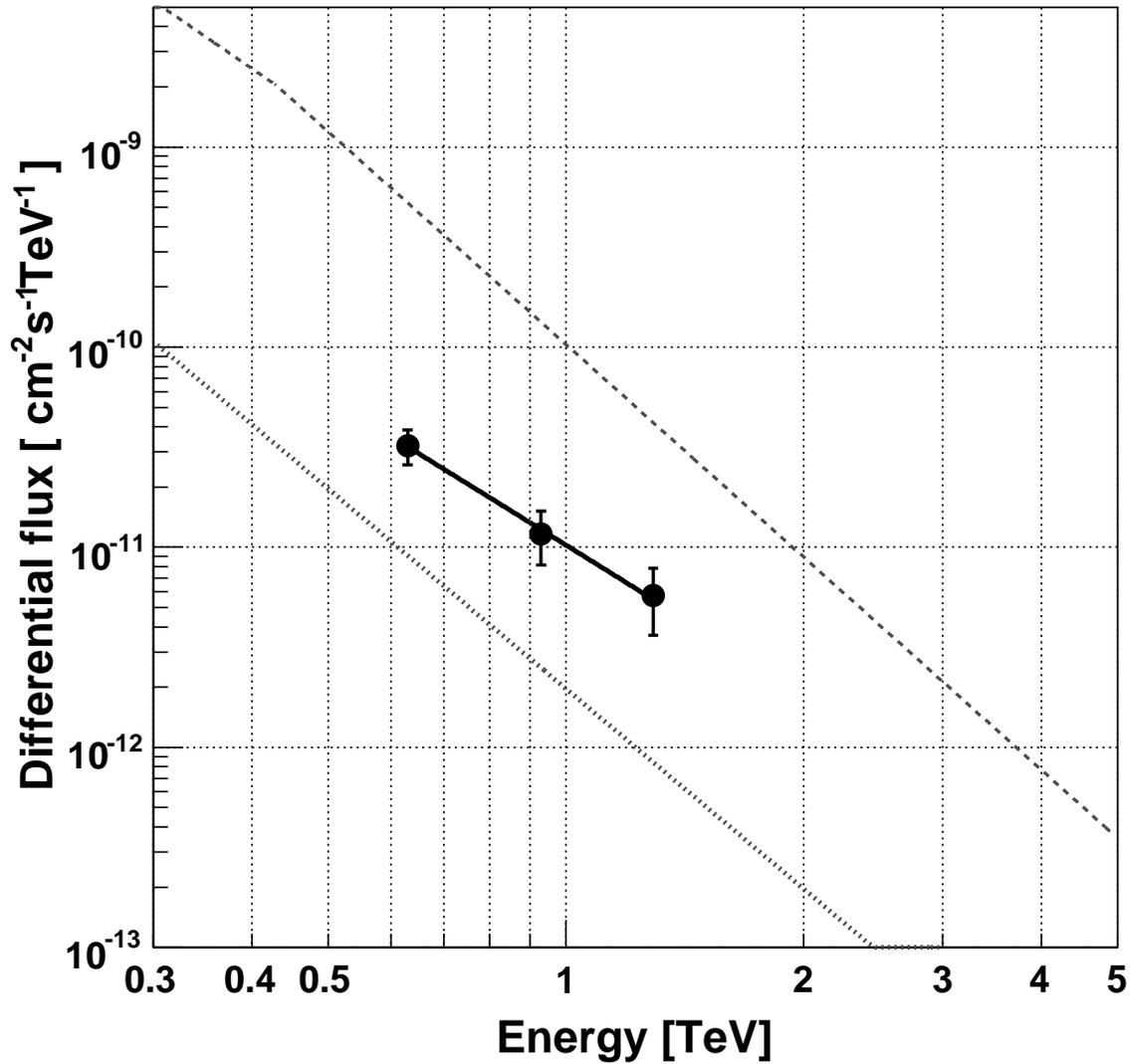}
\caption{Time-averaged differential energy spectrum 
of PKS\,2155$-$304 between July 28 and August 2. 
Data plotted by closed circles are obtained from all the data
at smaller zenith angle $z<30^{\circ}$ ($3fold$--$SZ$ and $2fold$--$SZ$). 
The solid line represents the best fit to these data 
assuming a single power law spectrum.
For comparison, the time average differential energy spectra
in early July 28 in 2007 reported by \citet{Aha07} 
during a giant flare and in 2002 and 2003 reported by \citet{Aha05a} 
at a low state are drawn by dashed and dotted lines, respectively.
\label{diff_spec}}
\end{figure}

We obtained the time-averaged differential energy spectrum
from all the data except for large zenith angle dataset,
{\it i.e.\/} from $3fold$--$SZ$ and $2fold$--$SZ$,
since the large zenith angle data have 
a higher energy threshold and 
therefore lower energy bins have a different exposure 
than higher energy one.
The differential flux is shown in Fig.~\ref{diff_spec} 
as closed circles with 1$\sigma$ statistical error bars.
The best fit of a power law to the small zenith angle data yields 
a photon index $\Gamma = 2.5\pm{0.5}_{stat}\pm{0.7}_{syst}$ 
and a flux normalization 
$N_0(~1~\mbox{TeV})~=~(1.0\pm0.2_{stat}\pm0.3_{syst})\times10^{-11}~\mbox{cm}^{-2}~\mbox{sec}^{-1}\mbox{TeV}^{-1}$.
We estimate the systematic error considering 
the same factors as mentioned in the previous section.

\subsection{Time variations}

\begin{figure}
\epsscale{}
\plotone{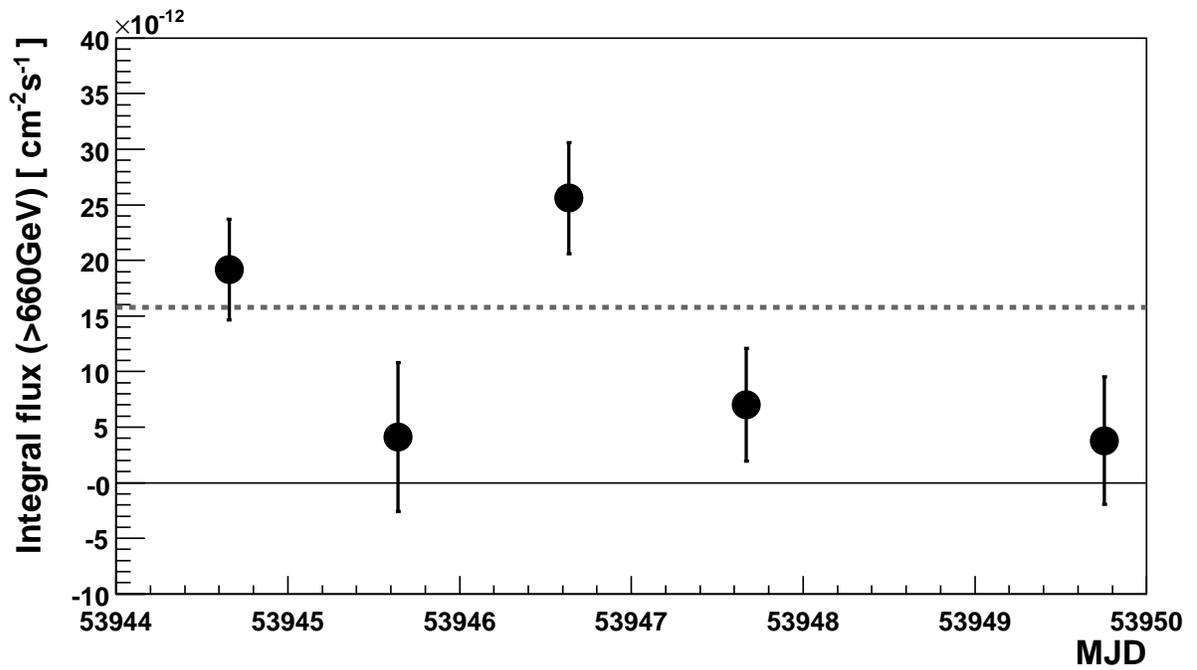}
\caption{Daily average integral flux of PKS\,2155$-$304 using 
all the data above 660\,GeV between 2006 July 28 and August 2.
Dashed line indicates an averaged integral flux during 
this observation period ($\sim45~\%$~Crab flux level).
\label{daybyday}}
\end{figure}

We have searched for gamma-ray emission from PKS\,2155$-$304
on a night-by-night basis.
The live time, $t_{liv}$, 
the number of on-source and off-source events, 
$N_{ON}$ and $N_{OFF}$,
the number of excess events, $N$, and 
significance, $s$, corresponding to each night are summarized 
in Table~\ref{summary}.
Here the off-source data are summed up over the five nights
to reduce the statistical fluctuation, 
and normalized to the live time of each night. 
The nightly average integral fluxes $F$ above 660\,GeV,
using all the data are also shown in the same table,
and are plotted in Fig.~\ref{daybyday},
where the flux from $3fold$ and $2fold$ datasets are combined.
The light curve shows that the average flux 
reached $\sim70~\%$~Crab in the night of July 30. 
%then 
%dropped rapidly by a factor of almost 3 during the same night
%on a time scale of $\sim2$~hours.
Assuming a constant average flux, a $\chi^2$ fit yields 
a value of 13.9 for 4 degrees of freedom, which corresponds to 
a $\chi^2$ probability of $\sim 0.8~\%$. 
This means that there is a marginal time variations 
of the average flux on a nightly basis between July 28 and August 2.  

\begin{figure}
\epsscale{}
\plotone{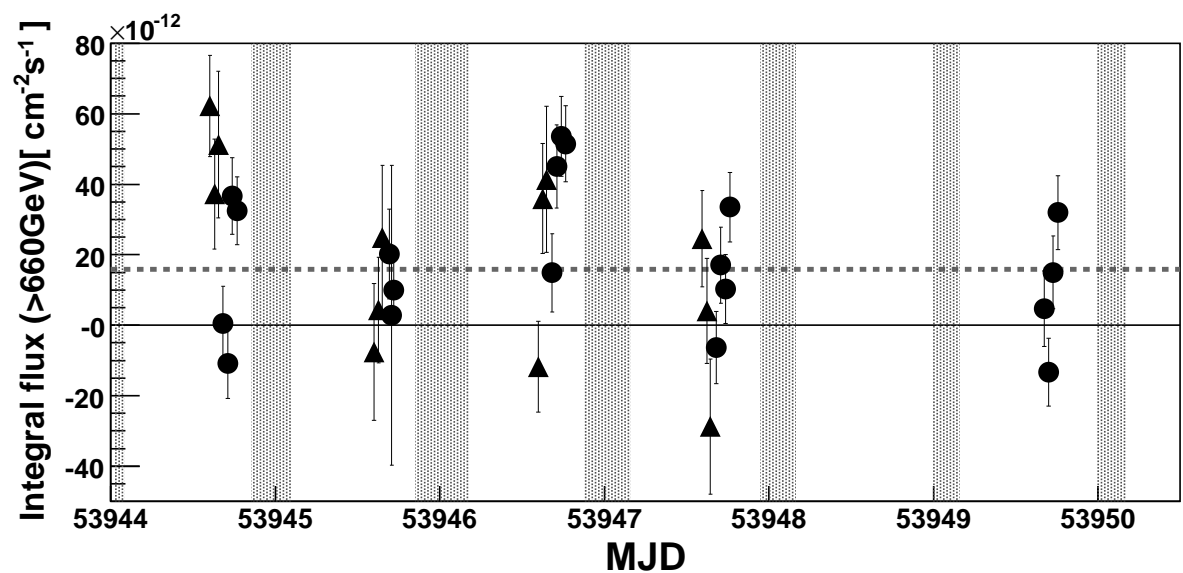}
\caption{Light curve of PKS\,2155$-$304 using all the data at zenith angle less than $30^{\circ}$ 
between 2006 July 28 and August 2, expressed by 
the integral flux above 660\,GeV. 
Closed triangles and closed circles indicate the results 
from $2fold$--$SZ$ and $3fold$--$SZ$ datasets, respectively. 
Dashed line indicates an average integral flux during this 
observation period.
The bin width is 40 minutes.
The shaded areas indicate the H.E.S.S.\ observation periods. 
\label{light_curve}}
\end{figure}

We further divide the data into $\sim40$ minutes bins 
for each night, and intra-night variation is investigated.
%Since the pointing position is alternatively shifted 
Since the pointing offset from the source position
is alternated in sign every 20 minutes 
in our observations, 40 minutes bin
is adopted as it offers a reasonable cancellation of 
any asymmetric effects.
The light curve of PKS\,2155$-$304 expressed 
by the integral flux of VHE gamma-rays above 660GeV 
using $2fold$--$SZ$ and $3fold$--$SZ$ dataset is
shown in Fig.~\ref{light_curve}.

Assuming a constant average flux for the data above 660\,GeV, 
a $\chi^2$ fit yields a value of 96.4 for 30 degrees of freedom, 
which corresponds to a $\chi^2$ probability of $7\times10^{-9}$. 
The same calculation was done for the data of July 28 and 30,
and gave $\chi^2$ values of 29.2 and 22.1 for 6 degrees of freedom, 
respectively.
The intra-night variations are apparent.
From figure~\ref{light_curve}, although it is difficult 
to calculate the fractional 
root mean square variability amplitude $F_{var}$ 
\citep{Vau03} because of the poor statistics,
$F_{var}=0.75\pm0.07$ on July 28 and $0.58\pm0.08$
on July 30 are obtained.

The results in the follow-up observations between August 17 and 25
are also summarized in Table~\ref{summary}.
The 2$\sigma$ upper limit above 680\,GeV is  
$F(>680\mbox{GeV})<6.8\times10^{-12}~\mbox{cm}^{-2}~\mbox{sec}^{-1} (<20\%~\mbox{Crab})$,
which shows TeV gamma-ray activity had subsided
compared to the outburst period.

%------------------------------------------------------------------------------
\section{Discussion}

From Figure~\ref{daybyday}, the fluxes on July 28 and 30 
exceed $50\%$ of the flux from the Crab nebula, 
and are the same level
as that at the end of July 27--28 (MJD 53944) observation 
reported by the H.E.S.S.\ group \citep{Aha07}, 
taking into account the difference of energy threshold 
and assuming a power law spectrum.
The flux in July 29 might be underestimated due to the presence
of clouds.
Therefore, our light curve suggests that the flux continuously
decreases on average following the large flare of July 28.
The preliminary H.E.S.S.\ light curve above 200\,GeV presented 
by \citet{Rau06} indicates that a 5~crab flare occurred
between our observations of July 29 and 30. However we did not
detect a flare in this period, possibly due to the 10~hour gap 
between the end of the H.E.S.S.\ observations and 
the start of our observations. 
For the shorter variability time scales, the July 28 and 30 data
indicate the intra-night variations.

The derived values of the fractional root mean square variability amplitude,
$F_{var}=0.49$ for night by night variation and 
$F_{var}=0.47$ for intra-night variation in July 28, 
are comparable to the values from the intra-night variability
earlier on July 28 reported by H.E.S.S. \citep{Aha07}.
They are rather higher than the values for the X-ray variability
(e.g., 0.10--0.43 listed by \citet{Zha99}),
(although much higher, and energy dependent, values
of $F_{var}$ have been reported by MAGIC~\citep{Alb07b} for Mrk~501).   

\begin{figure}
\epsscale{}
\plotone{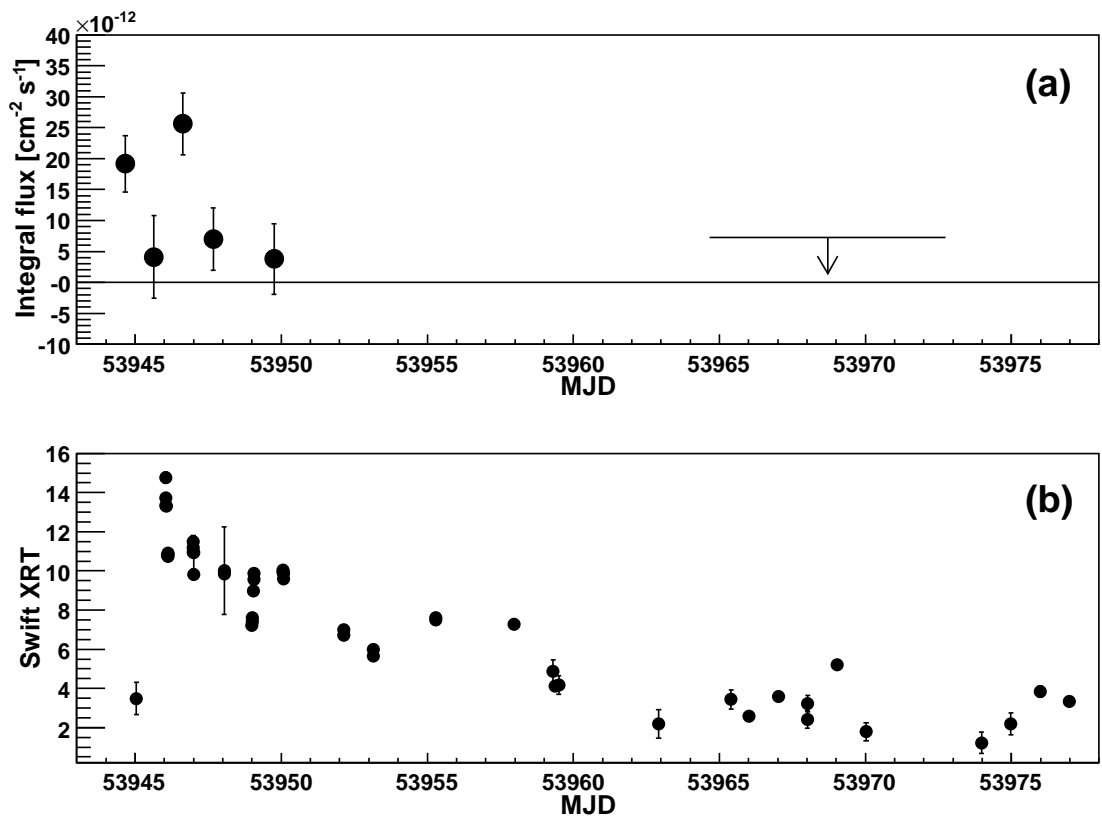}
\caption{Comparison of the light curves between VHE gamma-ray
and X-ray bands.(a) A nightly average integral flux 
of PKS\,2155$-$304 
above 660~GeV obtained by the CANGAROO-III 
given in unit of $\mbox{cm}^{-2}\mbox{s}^{-1}$, 
which is the same as in Fig.~\ref{daybyday}. 
An upper limit of the integral flux between August 17 and 25
is also plotted, which is an extrapolate value from the actual
upper limit above 680~GeV.
(b)~X-ray~[0.3--10~keV] counts in unit of $\mbox{count}~\mbox{s}^{-1}$
from the XRT on $Swift$ \citep{Fos07}.
\label{multiwave_lc}}
\end{figure}

In Figure~\ref{multiwave_lc}, our TeV light curve is
plotted together with the $Swift$ x-ray light curve~\citep{Fos07}.
Unfortunately, $Swift$ started observations two hours 
after CANGAROO stopped observing each day, so 
there are no simultaneous observations.
There is no evident correlation between the X-ray and the gamma-ray fluxes,
with a time lag of approximately 9~hours.
\citet{Fos07} reported the X-ray flux increased
by a factor of 5 in the 0.3--10\,keV energy band without
a large spectral change, and the highest peak 
of the $Swift$ light curve corresponds to
the second large TeV flare observed by H.E.S.S.~\citep{Rau06}.
The X-ray flux in this outburst is not very high and  
it is not possible to detect time variation of the flux
in the RXTE--ASM data.
For observations in 2002 and 2003, 
no correlation between the X-ray and the gamma-ray 
fluxes was reported by H.E.S.S.\ \citep{Aha05b}, however
a strong correlation during 2004 observations was presented 
in a preliminary analysis by \citet{Pun07}. 
On the other hand, Mrk~421 has displayed a good correlation
between the X-ray and the gamma-ray fluxes, although recently
no correlation results, including ``Orphan flares'', 
were reported.  
The simple one-zone SSC model is unable to explain 
such complex behavior ~\citep[e.g.,][]{Bla05}. 
Some models explaining ``Orphan flares'' have been proposed,
for example, by \citet{Kus06} for the leptonic model
and \citet{Bot05} for the hadronic model.

During the outburst periods, the photon index may change 
along with the flux variations, as is the case for Mrk~421 and Mrk~501
\citep[e.g.,][]{Aha02, Kre02, Alb07a, Alb07b}. 
However, the average photon index we obtained here 
does not differ from the value that H.E.S.S.\ reported 
previously \citep{Aha05a} within errors, 
and is consistent with those subsequently reported 
by H.E.S.S.\ \citep{Aha07} 
and MAGIC \citep{Maz07}.
The reason for this difference in the variation 
of photon index between PKS~2155$-$304 and Mrk~421 is 
still an open question.
Although the error on the spectrum is enormous, 
a spectral index after correction for the absorption
by the extra-galactic background light
is calculated to be $\Gamma = 1.6\pm0.5$ and $\Gamma = 1.3\pm0.5$
using the baseline model and the fast evolution model 
by \citet{Ste06}, respectively.
%Compared to the other blazars, these values are moderate
%\citep{Ste07}. 

%------------------------------------------------------------------------------
\section{Conclusion}

We observed the southern HBL PKS~2155$-$304 from
2006 July 28 (MJD 53944) to August 2 with the CANGAROO-III imaging 
Cherenkov telescopes. During the VHE high state
we detected a signal at $4.8\sigma$ significance 
above 660~GeV 
in the total effective exposure time of 25.1~hours.
The time averaged integral flux above 660~GeV is
$(1.6\pm0.3_{stat}\pm0.5_{syst})\times10^{-11}\,\mbox{cm}^{-2}\,\mbox{sec}^{-1}$
which corresponds to $\sim45\%$ of the flux observed 
from the Crab nebula. 
The intra-night time variations of the flux 
were    seen in our observations.
Follow-up observations two weeks later
indicated the source activity had decreased to lower than 
$\sim20\%$ Crab flux..

%------------------------------------------------------------------------------
\acknowledgments

We thank Dr.\ W.\ Hofmann, Dr.\ S.\ Wagner, Dr.\ G.\ Rowell, 
Dr.\ W.\ Benbow, Dr.\ B.\ Giebels, and L.\ Foschini 
for providing details of the H.E.S.S. and the Swift 
observations of PKS~2155$-$304. 
This work is supported by a Grant-in-Aid for Scientific Research by the
Japan Ministry for Education, Culture, Sports, Science and Technology,
the Australian Research Council (Grants LE0238884 and DP0345983), 
and Inter-University Research Program by the Institute for Cosmic Ray Research.
The support of JSPS Research Fellowship for T.N.\ and Y.H.\ is
gratefully acknowledged.
A part of this work was funded for Y.S. by the Sasagawa Scientific 
Research Grant from The Japan Science Society.

%------------------------------------------------------------------------------

\clearpage

%------------------------------------------------------------------------------
\begin{deluxetable}{ccccccc}
\tablewidth{0pt}
\tabletypesize{\small}
\tablecaption{Summary of observations for PKS\,2155$-$304 from 2006 July
 28 (MJD 53944) to August 2 (MJD 53944) and the follow-up observations 
from August 17 (MJD 53964) to 25 (MJD 53972)
\tablenotemark{a}
\label{mjd}}
\tablehead{
\colhead{} & 
\colhead{} & 
\colhead{Begin} & 
\colhead{End} &
\colhead{$t_{obs}$} & 
\colhead{$z$} &
\colhead{$r_{tr}$} \\
\colhead{Date} & 
\colhead{$N_{tel}$} & 
\colhead{[MJD]} & 
\colhead{[MJD]} & 
\colhead{[hrs]} & 
\colhead{[${}^{\circ}$]} &
\colhead{[Hz]} 
%\colhead{[Hz]}
}
\startdata
July 28 & 2 & 53944.514 & 53944.663 & 3.6 & 29.8 & 18.7 \\
& 3 & 53944.668 & 53944.830 & 3.9 & 20.4 & 12.1 \\
July 29 & 2 & 53945.522 & 53945.660 & 3.1 & 27.6 & 14.9 \\
& 3 & 53945.679 & 53945.762 & 2.0 & 12.1 &  6.2 \\
July 30 & 2 & 53946.535 & 53946.656 & 2.9 & 26.1 & 20.1 \\
& 3 & 53946.668 & 53946.833 & 4.0 & 22.2 & 12.5 \\
July 31 & 2 & 53947.576 & 53947.653 & 1.8 & 19.5 & 19.8 \\
& 3 & 53947.666 & 53947.827 & 3.9 & 21.7 & 11.6 \\
Aug.  2 & 3 & 53949.658 & 53949.822 & 3.9 & 21.5 & 11.9 \\
July 28--Aug.2 & 2 & & & 11.4 & 26.6 & 18.2 \\
& 3 & & & 17.6 & 20.4 & 11.4 \\
Aug. 17--25 & 3 & & & 19.1 & 20.9 & 10.9 \\
\enddata
\tablenotetext{a}{Observation date, the number of used telescopes, 
$N_{tel}$, begin and end time of each observation in MJD, 
are summarized from the first column to the fourth column. 
In the following three columns,
observation time, $t_{obs}$, average zenith angle, $z$, and 
average trigger rate, $r_{tr}$ are shown. 
Combined data between July 28 and August 2 are also shown.
For the follow-up observations, only combined data are summarized.}
\end{deluxetable}
\clearpage

%------------------------------------------------------------------------
\begin{deluxetable}{ccccccccc}
\tablewidth{0pt}
\tabletypesize{\small}
\tablecaption{Summary of results for PKS\,2155$-$304 from 2006 July
 28 (MJD 53944) to August 2 (MJD 53944) and the follow-up observations 
from August 17 (MJD 53964) to 25 (MJD 53972)
\tablenotemark{a}
\label{summary}}
\tablehead{
\colhead{Date} & 
\colhead{$N_{tel}$} & 
\colhead{$r_{sh}$[Hz]} & 
\colhead{$t_{liv}$[hrs]} & 
\colhead{$N_{ON}$} &
\colhead{$N_{OFF}$} &  
\colhead{$N$} & 
\colhead{$s$[$\sigma$]} &
\colhead{$F$($>$660GeV)}\tablenotemark{b} 
}
\startdata
July 28 & 2 & 12.4 & 3.4 & 675 & 545 & $130\pm29$ & 4.6 & $3.3\pm0.7$ \\
& 3 & 8.0 &3.5 & 157 & 132 &  $25\pm14$ & 1.8 & $1.0\pm0.6$ \\
July 29\tablenotemark{c} & 2 & 10.2 &2.2 & 338 & 340 &   $-2.4\pm20$ & -0.12 & $-0.10\pm0.8$\\
& 3 & 4.1 &0.9 & 39 & 32 &  $7.1\pm6.1$ & 1.2 & $1.3\pm1.1$ \\
July 30 & 2 & 13.2 &2.8 & 434 & 423 & $11\pm23$ & 0.48 & $0.31\pm1.0$ \\
& 3 & 8.1 & 3.5 & 235 & 135 & $100\pm16$ & 6.1 & $3.9\pm0.6$ \\
July 31 & 2 & 13.0 & 1.7 & 229 & 224 & $4.8\pm17$ & 0.29 & $2.9\pm1.0$ \\
& 3 & 7.6 & 3.6 & 156 & 135 & $21\pm15$ & 1.4 & $0.81\pm0.57$ \\
Aug.  2 & 3 & 7.8 & 3.5 & 142 & 133 & $9.4\pm14$ & 0.66 & $0.38\pm0.57$ \\ 
July 28--Aug.2 & 2 & 12.1 & 10.1 &1676 & 1532 & $144\pm56$ & 2.5 & $1.3\pm0.5$ \\
& 3 & 7.4 & 15.0 & 729 & 551 &$178\pm36$ & 5.0 & $1.7\pm0.3$ \\ \hline
Aug. 17--25 & 3 & 7.4 & 17.1 & 625 & 637 &$-12\pm36$ & $-0.34$ & $-0.12\pm0.34$\tablenotemark{d} 
\enddata
\tablenotetext{a}{Following the observation date and the number of 
telescopes, average shower rate, $r_{sh}$, and live time, $t_{liv}$,
are shown in the third and fourth columns, respectively.
$N_{ON}$ and $N_{OFF}$ in the fifth and sixth columns are
the number of on-source and off-source events, respectively,
where the off-source data from July 28 through August 2
were summed up for five nights and normalized to the live time.
Seventh column indicates the number of excess events derived 
by subtraction of off-source events from on-source events,
and their significance are shown in the eighth column. 
Integral flux above 660\,GeV are calculated and are shown 
in the last column. Combined results from July 27 to August 2
and from August 17 and 25 are also presented, respectively.}
\tablenotetext{b}{in unit of $\times10^{-11}\,\mbox{cm}^{-2}\,\mbox{sec}^{-1}$}
\tablenotetext{c}{affected by clouds}
\tablenotetext{d}{Threshold energy is 680\,GeV.}
\end{deluxetable}
\clearpage
%------------------------------------------------------------------------

\clearpage

\clearpage

\clearpage

\clearpage

\clearpage

\clearpage


\begin{thebibliography}{}
\bibitem[Aharonian et al.(1999)]{Aha99}
    Aharonian, F., et al. 1999, \aap, 342, 69
\bibitem[Aharonian et al.(2002)]{Aha02}
    Aharonian, F., et al. 2002, \aap, 393, 89
\bibitem[Aharonian et al.(2004)]{Aha04}
    Aharonian, F., et al. 2004, \apj, 614, 897
\bibitem[Aharonian et al.(2005a)]{Aha05a}
    Aharonian, F., et al. 2005a, \aap, 430, 865
\bibitem[Aharonian et al.(2005b)]{Aha05b}
    Aharonian, F., et al. 2005b, \aap, 442, 895
\bibitem[Aharonian et al.(2005c)]{Aha05c}
    Aharonian, F., et al. 2005c, \aap, 437, 95
\bibitem[Aharonian et al.(2007)]{Aha07}
    Aharonian, F., et al. 2007, \apj, 664, L71 
\bibitem[Albert et al.(2007a)]{Alb07a}
    Albert, J., et al. 2007, \apj, 663, 125
\bibitem[Albert et al.(2007b)]{Alb07b}
    Albert, J., et al. 2007, \apj, submitted (astro-ph/0702008)
\bibitem[Beilicke(2006)]{Bei06}
    Beilicke, M. 2006,
    presentation at the 3rd Workshop on the Nature of Unidentified 
    High-Energy Sources (Barcelona)
\bibitem[Benbow et al.(2006)]{Ben06}
    Benbow, W., Costamante, L., \& Giebels, B. 2006, ATel \#867
\bibitem[Blazejowski et al.(2005)]{Bla05}
    Blazejowski, M., et al. 2005, \apj, 630, 130
\bibitem[B\"ottcher (2005)]{Bot05}
    B\"ottcher, M. 2005, \apj, 621, 176
\bibitem[Brinkmann et al.(1994)]{Bri94}
    Brinkmann, W., et al. 1994, \aap, 288, 443
\bibitem[Buckley et al.(1996)]{Buc96}
    Buckley, J. H., et al. 1996, \apj, 472, L9
\bibitem[Chadwick et al.(1999a)]{Cha99a}
    Chadwick, P. M., et al. 1999a, \apj, 513, 161
\bibitem[Chadwick et al.(1999b)]{Cha99b}
    Chadwick, P. M., et al. 1999b, 
    in Proc.\ 26th Int.\ Cosmic Ray Conf.\ (Salt Lake City), 3, 338
\bibitem[Chiappetti et al.(1999)]{Chi99}
    Chiappetti, L., et al. 1999, \apj, 521, 552
\bibitem[Cui (2004)]{Cui04}
    Cui, W. 2004, \apj, 605, 662
\bibitem[Dermer \& Schlickheiser(1993)]{Der93}
    Dermer, C. D., \& Schlickheiser, R. 1993, \apj, 416, 458
\bibitem[Djannati-Atai et al.(1999)]{Dja99}
    Djannati-Atai, A., et al. 1999, \aap, 350, 17
\bibitem[Dominici et al.(2004)]{Dom04}
    Dominici, T. P., et al. 2004, \aj, 128, 47
\bibitem[Edelson et al.(1995)]{Ede95}
    Edelson, R. A., et al. 1995, \apj, 438, 120
\bibitem[Edelson et al.(2001)]{Ede01}
    Edelson, R. A., et al. 2001, \apj, 554, 274
\bibitem[Enomoto et al.(2006a)]{Eno06a}
    Enomoto, R., et al. 2006a, \apj, 638, 397
\bibitem[Enomoto et al.(2006b)]{Eno06b}
    Enomoto, R., et al. 2006b, \apj, 652, 1268
\bibitem[Falomo et al.(1993)]{Fal93}
    Falomo, R. et al. 1993, \apj, 411, L63
\bibitem[Fisher(1936)]{Fis36}
    Fisher, R. A. 1936, Annals of Eugenics, 7, 179
\bibitem[Foschini et al.(2007)]{Fos07}
    Foschini, L., et al. 2007, \apj, in press(astro-ph/0701868)
\bibitem[Gaidos et al.(1996)]{Gai96}
    Gaidos, J. A., et al. 1996, Nature, 383, 319
\bibitem[Giebels(2006)]{Gie06}
    Giebels, B. 2006, 
    presentation at 2nd Workshop on TeV Particle Astrophysics (Madison)
\bibitem[Giommi et al.(1998)]{Gio98}
    Giommi, P., et al. 1998, \aap, 333, L5
\bibitem[Gliozzi et al.(2006)]{Gli06}
    Gliozzi, M., et al. 2006, \apj, 646, 61
\bibitem[Griffiths et al.(1979)]{Gri79}
    Griffiths, R. E., et al. 1979, \apj, 234, 810
\bibitem[Hayashida et al.(1998)]{Hay98}
    Hayashida, K., et al. 1998, \apj, 500, 642
\bibitem[Hillas et al.(1985)]{Hil85}
    Hillas, A. M., et al. 1985
    in Proc.\ 19th Int.\ Cosmic Ray Conf.\ (La Jolla), 3, 445
\bibitem[Hinton (2007)]{Hin07}
    Hinton, J. 2007, 
    in Proc.\ of the 30th Int.\ Cosmic Ray Conf.\ (Merida), Rapporteur Talk
\bibitem[Hofmann et al.(1999)]{Hof99}
    Hofmann, W. et al. 1999, Astropart. Phys., 122, 135
\bibitem[Jones, O'dell \& Stein(1974)]{Jon74}
    Jones, T. W., O'dell, S. L., \& Stein, W. A. 1974, \apj, 188, 353
\bibitem[Kabuki et al.(2003)]{Kab03}
    Kabuki, S., et al. 2003, NIM, A500, 318
\bibitem[Kabuki et al.(2007)]{Kab07}
    Kabuki, S., et al. 2007, \apj, 668, 968
\bibitem[Kataoka et al.(2000)]{Kat00}
    Kataoka, J., et al. 2000, \apj, 528, 243
\bibitem[Kataoka et al.(2001)]{Kat01}
    Kataoka, J., et al. 2001, \apj, 560, 659
\bibitem[Kawachi et al.(2001)]{Kaw01}
    Kawachi, A., et al. 2001, Astropart. Phys., 14, 261
\bibitem[Krawczynski et al.(2001)]{Kra01}
    Krawczynski, H., et al. 2001, \apj, 559, 187
\bibitem[Krawczynski et al.(2002)]{Kra02}
    Krawczynski, H., et al. 2002, MNRAS, 336, 721
\bibitem[Krawczynski et al.(2004)]{Kra04}
    Krawczynski, H., et al. 2004, \apj, 601, 151
\bibitem[Krennrich et al.(2002)]{Kre02}
    Krennrich, F., et al. 2002, \apj, 575, L9
\bibitem[Kubo et al.(1998)]{Kub98}
    Kubo, H., et al. 1998, \apj, 504, 693
\bibitem[Kubo et al.(2001)]{Kub01}
    Kubo, H., et al. 2001, 
    in Proc.\ 27th Int.\ Cosmic Ray Conf.\ (Hamburg), 2900
\bibitem[Kusunose \& Takahara(2006)]{Kus06}
    Kusunose, M., \& Takahara, F. 2006, \apj, 651, 113
\bibitem[Marshall, Carone, \& Fruscione(1993)]{Mar93}
    Marshall, H. L., Carone, T. E., \& Fruscione, A. 1993, \apj, 414, L53 
\bibitem[Mazin \& Lindfors(2007)]{Maz07}
    Mazin, D., \& Lindfors, E. 2007, 
    in Proc.\ of the 30th Int.\ Cosmic Ray Conf.\ (Merida), OG2.3, \#936
\bibitem[Nakase et al.(2003)]{Nak03}
    Nakase, T. 2003,
    in Proc.\ 28th Int.\ Cosmic Ray Conf.\ (Tsukuba), 2587
\bibitem[Nicastro(2002)]{Nic02}
    Nicastro, F. 2002, \apj, 573, 157
\bibitem[Nishijima(2002)]{Nis02}
    Nishijima, K. 2002, PASA, 19, 26
\bibitem[Nishijima et al.(2001)]{Nis01}
    Nishijima, K., et al. 2001,
    in Proc.\ 27th Int.\ Cosmic Ray Conf.\ (Hamburg), 2626
\bibitem[Nishijima et al.(2005)]{Nis05}
    Nishijima, K., et al. 2005,
    in Proc.\ 29th Int.\ Cosmic Ray Conf.\ (Pune), 15, 327
\bibitem[Padovani \& Giommi(1995)]{Pad95}
    Padovani, P., \& Giommi, P. 1995, \apj, 444, 567
\bibitem[Pian et al.(1997)]{Pia97}
    Pian, E., et al. 1997, \apj, 486, 784
\bibitem[Piner \& Edwards(2004)]{Pin04}
    Piner, E., \& Edwards, P. G. 2004, \apj, 600, 115
\bibitem[Punch et al.(1992)]{Pun92}
    Punch, M., et al. 1992, Nature, 358, 477
\bibitem[Punch et al.(2007)]{Pun07}
    Punch, M., et al. 2007, presented at the 27th Int.\ Cosmic Ray
				Conf.\ (Merida)
\bibitem[Quinn et al.(1996)]{Qui96}
    Quinn, J., et al. 1996, \apj, 456, L83 
\bibitem[Raue et al.(2006)]{Rau06}
    Raue, M., et al. 2006, presented at the INTEGRAL Workshop on the keV
				to TeV Connection(Rome)
\bibitem[Rebillot et al.(2006)]{Reb06}
    Rebillot, P. F., et al. 2006, \apj, 641, 740
\bibitem[Roberts et al.(1999)]{Rob99}
    Roberts, M. D., et al. 1999, \aap, 343, 691
\bibitem[Sambruna et al.(2000)]{Sam00}
    Sambruna, R. M., et al. 2000, \apj, 538, 127
\bibitem[Schwartz et al.(1979)]{Sch79}
    Schwartz, D. A., et al. 1979, \apj, 229, L53
\bibitem[Sikora, Begelman \& Rees(1994)]{Sik94}
    Sikora, M., Begelman, M. C., \& Rees, M. J. 1994, \apj, 421, 153
\bibitem[Sreekumar \& Vestrand(1997)]{Sre97}
    Sreekumar, P., \& Vestrand, W. T. 1997, IAU Circular 6776
%\bibitem[Stecker et al.(2007)]{Ste07}
%    Stecker, F. W., et al. 2007, \apj, submitted
\bibitem[Stecker \& Scully(2006)]{Ste06}
    Stecker, F. W., \& Scully, S. T. 2006, \apj, 652, L9
\bibitem[Takahashi et al.(1996)]{Tak96}
    Takahashi, T., et al. 1996, \apj, 470, L89
\bibitem[Takahashi et al.(2000)]{Tak00}
    Takahashi, T., et al. 2000, \apj, 542, L105
\bibitem[Tanihata et al.(2001)]{Tan01}
    Tanihata, C., et al. 2001, \apj, 563, 569
\bibitem[Treves et al.(1989)]{Tre89}
    Treves A., et al. 1989, \apj, 341, 733
\bibitem[Ulrich, Maraschi, \& Urry(1997)]{Ulr97}
    Ulrich, M.-H., Maraschi, L., \& Urry, C. M., 1997, \araa, 35, 445
\bibitem[Urry et al.(1997)]{Urr97}
    Urry, C. M., et al. 1997, \apj, 486, 799
\bibitem[Vaughan et al.(2003)]{Vau03}
    Vaughan, S., et al. 2003, MNRAS, 345, 1271
\bibitem[Vestrand \& Sreekumar(1999)]{Ves99}
    Vestrand, W. T., \& Sreekumar, P. 1999, Astropart.\ Phys., 11, 197
\bibitem[Vestrand, Stacy, \& Sreekumar(1995)]{Ves95}
    Vestrand, W. T., Stacy, J. G., \& Sreekumar, P. 1995, \apj, 454, L93
\bibitem[Zhang et al.(1999)]{Zha99}
    Zhang, Y. H., et al. 1999, \apj, 527, 719
\bibitem[Zhang et al.(2002)]{Zha02}
    Zhang, Y. H., et al. 2002, \apj, 572, 762
\bibitem[Zhang et al.(2005)]{Zha05}
    Zhang, Y. H., et al. 2005, \apj, 629, 686
\bibitem[Zhang et al.(2006)]{Zha06}
    Zhang, Y. H., et al. 2006, \apj, 651, 782
\bibitem[Zhang \& Xie(1996)]{Zha96}
    Zhang, Y. H., \& Xie, G. Z. 1996, \aap, 116, 289

\end{thebibliography}
\end{document}